# QUANTITATIVE EVALUATION OF THE HYPOTHESIS THAT BL LACERTAE OBJECTS ARE QSO REMNANTS

E.F. Borra

Centre d'Optique, Photonique et Lasers
Département de Physique, Université Laval

# ABSTRACT

We evaluate with numerical simulations the hypothesis that BL Lacertae objects are the remnants of Quasi Stellar Objects. This hypothesis is based on their highly peculiar redshift evolution. They have a comoving space density that increases with decreasing redshift, contrary to all other active galactic nuclei. We assume that relativistic jets are below detection in young radio-quiet quasars and increase in strength with cosmic time so that they eventually are detected as BL Lacertae objects. Our numerical simulations fit very well the observed redshift distributions of BL Lacertae objects. There are strong indications that only the high-synchrotron-peaked BL Lacertae objects could be QSO remnants.

# 1. INTRODUCTION

The BL Lacertae objects (BLLs) are a subgroup of Active Galactic Nuclei (AGNs) generated by relativistic beaming (Blandford & Rees 1978). Unlike other AGNs, they show negative evolution with redshift, which implies an increase in comoving space density with cosmic time. However, this evidence has been both confirmed and disputed by a series of papers (Ajello et al. 2014). A weakness of these papers comes from the small numbers of observed objects. It is less important in Ajello et al. (2014) who discuss the cosmic evolution of BLLs using a sample of 211 γ-ray selected BLLs. They find that the situation is complex. The spectral energy distributions of BLLs extend over a very large frequency range ( $10^8 < \nu < 10^{27}$ Hz) as can be seen in figure 2 in Falomo, Pian & Treves(2014). The low frequency component ($\nu < 10^{18}$ Hz) is caused by synchrotron radiation. Ajello et al. (2014) divided the objects into three classes separated according to the peak of the synchrotron component: high-synchrotron-peaked (HSP, $\nu_{peak} > 10^{15}$ Hz), intermediate-synchrotron-peaked (ISP, $10^{14} < \nu_{peak} < 10^{15}$ Hz) and low-synchrotron-peaked (LSP, $\nu_{peak} < 10^{14}$ Hz). They find that the negative evolution is entirely driven by low luminosity HSPs. All ISPs and LSPs show positive evolution.

Although a cosmological test that uses the knowledge of the space density of quasar remnants at $z = 0.0$ was the main subject in Borra (1983), it made the novel hypothesis that BLLs are beamed remnants of Quasi Stellar Objects (QSOs) to obtain the space density of QSO remnants at $z = 0.0$. The fact that it explains why BLLs do not undergo the strong positive evolution seen in other AGNs but instead have negative evolution was the reason for this hypothesis. Borra (1983) assumed that all Quasi Stellar Objects (QSOs), radio-quiet (RQQ) as well as radio-loud (RLQ), are capable of generating relativistic jets but that jets are snuffed in young RQQs to later emerge in aged RQQs. A QSO remnant is detected as a BLL object only if the angle between the jet and the line of sight is small.

In this paper, we reexamine the hypothesis that BLLs are beamed remnants of QSOs on the basis of three decades of progress in studies of BLLs.

## 2. NUMERICAL MODELLING

We start from the assumption that all quasars, radio quiet as well as radio loud, are capable of generating jets but that the jets are below detection in young radio-quiet quasars and that they emerge later in cosmic time. We do not make any assumption for the physical reasons for this hypothesis. This is not needed because the fact that the commoving space density of BLLs increases with decreasing redshift clearly shows that, whatever the physics, the relativistic jets emerge later in cosmic time.

Our numerical models start from a luminosity function, modeled by a power law (Equation 2), for QSO remnants. Urry & Shafer (1984) start from this luminosity function and evaluate the effects on it from relativistic jets in the core of AGNs. The relativistic jet has a beaming effect that greatly increases the luminosity within a narrow angle along the direction of the jet. This has a considerable effect on the luminosity function since it greatly increases the observed luminosity only for the very small fraction of AGNs which send the beam within a narrow range of angles in the direction of the observer. Urry & Shafer (1984) derive an equation for this luminosity function (Equation 3). Although, at first sight, Equation 3 appears to depend on several parameters, in practice, besides the parameters from the unbeamed luminosity function, it only depends on the relativistic Lorentz factor $\gamma$, the range of angles $\theta$ in the beam under consideration and the parameter $p$ in Equation 1 that relates the observed luminosity of a relativistic jet to its emitted luminosity. The numerical redshift distribution that we then compare to the observation is obtained from the usual cosmology relation (Equation 5), where we use the beamed luminosity function (Equation 3) multiplied by $(1+z)^b$ to quantify the evolution with cosmic time of the luminosity function (Equation 6). Equation 3 has the very important normalization parameter $K$ which is obtained from Equation 4 that relates the unbeamed QSO remnant luminosity function $\Phi_{QR}(M,z=0)$ at $z=0$ to the quasar luminosity function $\Phi_{QSO}(M,z=2.0)$ at $z=2.0$. We obtain $\Phi_{QSO}(M,z=2.0)$ from Ross et al. (2012). The next paragraph briefly summarizes how Equation 3 is derived. The other paragraphs in this section describe how we evaluate the parameters in Equation 3.

Urry & Shafer (1984) start from the equation that relates the observed luminosity $L$ of a relativistic jet to its emitted luminosity $\mathscr{L}$

$$L = \delta^p \mathscr{L} \quad , \quad (1)$$

where $\delta = [\gamma(1-\beta\cos\theta)]^{-1}$, with $\gamma$ the Lorentz factor, $\beta$ the velocity in units of the speed of light and $\theta$ the angle between the velocity vector and the line of sight. They assume that the isotropic unbeamed component of the luminosity function has the power law

$$\Phi(\mathscr{L}) = K \mathscr{L}^{-B} \quad \text{for} \quad \mathscr{L}_1 < L < \mathscr{L}_2$$
$$\Phi(\mathscr{L}) = 0 \quad \text{for} \quad \mathscr{L} < \mathscr{L}_1 \text{ or } \mathscr{L} > \mathscr{L}_2 \quad . \quad (2)$$

The upper and lower luminosity cutoffs approximate the fact the luminosity function must turn down at the luminosity extremes (Urry & Shafer 1984). They find that relativistic beaming converts the power law luminosity function in Equation (2) to the luminosity function given by

$$\Phi(L) = 0 \quad \text{for } L < L_{min}$$

$$\Phi(L) = K/(\beta\gamma pC)\, \mathcal{L}_1^{-C} L^{-(p+1)/p} \quad \text{for} \quad L_{min} < L < L_4 \qquad (3)$$

$$\Phi(L) = K/(\beta\gamma pC)\, \delta_{max}^{pC} L^{-B} \quad \text{for} \quad L_4 < L < L_{max}$$

$$\Phi(L) = 0 \quad \text{for } L > L_{max}$$

with $C = B - (1/p) - 1$ and $\delta_{max} = [\gamma(1-\beta\cos\theta_{min})]^{-1}$ $L_{min} = \delta_{min}^{p}\, \mathcal{L}_1$, $\delta_{min} = [\gamma(1-\beta\cos\theta_{max})]^{-1}$, $L_4 = \delta_{max}^{p}\, \mathcal{L}_1$, $L_{max} = \delta_{max}^{p}\, \mathcal{L}_2$ where $\theta$ is the angle of the jet with respect to the line of sight and $\theta_{min}$ and $\theta_{max}$ are respectively the lower and upper limits of the range of angles $\theta$ under consideration. The high luminosity end has an exponential index $-B$, and the low luminosity end has an exponential index $-(p+1)/p$, where $p$ is in the range $3 < p < 5$ and depends on the spectrum, the structure of the jet and the frequencies being compared.

Ross et al. (2012) present measurements of the optical Quasar luminosity functions using data from the Sloan Digital Sky Survey. Figure 11 in Ross et al.(2012) gives the luminosity functions in 8 redshifts intervals ranging from $z = 0.3$ to $z = 3.5$. It shows that the number density peaks at $z \sim 2.0$ and decreases by over of factor of 10 to $z = 0.2$. We can therefore make the reasonable assumption that the density of QSO remnants at $z = 0.0$ should be approximately equal to the density of quasars at $z = 2$ and therefore

$$\int_{M_{lr}}^{M_{ur}} \Phi_{QR}(M, z=0.0)\, dM = \int_{M_{lq}}^{M_{uq}} \Phi_{QSO}(M, z=2.0)\, dM \quad , \qquad (4)$$

where $\Phi_{QR}(M,z)$ and $\Phi_{QSO}(M,z)$ are respectively the luminosity functions of QSO remnants and QSOs. Equation (4) assumes that all QSOs are created in a narrow redshift range centered at $z= 2.0$. This obviously approximates the actual situation for they are probably continuously created as a function of $z$; however figure 11 in Ross et al. (2012) clearly shows that the majority are created around $z = 2$, consequently this approximation is adequate. The $M_{ur}$ and $M_{lr}$ for the QSO remnants, and $M_{uq}$ and $M_{lq}$ for the QSOs, upper and lower magnitude limits in Equation (4) simply model the fact that the luminosity functions turn down at some upper and lower magnitudes. We shall now carry out numerical simulations using the luminosity function in Equation (3) subject to the condition set by Equation (4).

The redshift distribution of BLLs given by figure 9 in Plotkin et al.(2008) shows a peak at $z \sim 0.3$ followed by a rapid decrease for $z > 0.6$. This is consistent with a sharply peaked luminosity function combined with a negative evolution of the space density of the BLLs. Equation (3) generates such a luminosity function as can be seen in figure 2 in Urry & Shafer (1984). Note however that the luminosity functions in Urry & Shafer (1984) are multiplied by the luminosity $L$, so that the luminosity function alone is actually much strongly peaked than in figure 2 in Urry & Shafer (1984). We obtain a good fit to the redshift distributions in figure 9 in Plotkin et al.(2008) with an unbeamed power-law luminosity function in the magnitude range $-11 < M > -13$ having an index $B= 4.0$. The unbeamed luminosity function is restricted by the fact that the local QSO remnants are undetected and therefore must be much less luminous than QSOs. The $M_{ur} = -11$ and $M_{lr} = -13$ range for the QSO remnants in Equation 4 respect this criterion. This narrow magnitude range comes from the steep power law of the unbeamed luminosity function (Equation 2) so that there is little contribution for faint objects. This luminosity function respects the requirement that the unbeamed component be inconspicuous. We obtain the normalization factor $K$ from Equation (4), with the luminosity function $\Phi_{QSO}(M,z)$ from Ross et al. (2012) The integral of the luminosity function $\Phi_{QSO}(M,z=2)$ in Equation 4 is obtained from the total number of quasars in the redshift range *1.82 < z < 2.20* in figure 11 in Ross et al. (2012). We use the lower limit $M_{lq}$=-22 by assuming that QSOs fainter than *M* = -22 do not evolve enough to contribute significantly to the local population of *QSO* remnants. This assumption comes from Figure 11 in Ross et al. (2012) which shows that the comoving space density of QSOs at $M_i$ = *-22* changes little with redshift, The upper limit $M_{uq}$=-30 comes from Figure 11 in Ross et al. (2012) which shows a negligible space density of QSOs with $M_i$ < *- 30* at *z = 2.0*. There are obvious uncertainties in our estimate of *K*, coming from the determination of the luminosity function of QSOs, from uncertainties in the cosmological parameters and from the lower magnitude cutoff. We use a relativistic Lorentz factor $\gamma$ = 6.1, which is the $\gamma$ value obtained for the BLLs by Ajello et al. (2014) and *p*= 4 for the exponent in Equation (3), a value which is in agreement with Urry & Shafer (1984) since we use broadband luminosity. Note also that figure 8 in Ajello et al, (2014) that compares the beaming model to the observations also uses *p* = 4. Figure 8 in Ajello et al. (2014) shows the average luminosity functions of BLLs, confirming that the exponential luminosity function assumed by Urry & Shafer (1984) (Equation 2) is a good approximation.

The beamed luminosity function is computed within viewing angles $\theta$ in the range 0.0º < $\theta$ < 7º . The upper limit of *7º* was chosen to include *2/3* of the viewing angles, in the distribution of viewing angles in figure 9 in Ajello et al. (2014) , because it sets the luminosity at which the luminosity function turns down. Including all of the angles would have given too much weight to angles higher than 7°. Note also that increasing the upper limit by a few degrees does not affect significantly the shape of the theoretical distribution in Figure 1 because it only increases the turn-down luminosity of the very steep luminosity function given by Equation 3. The redshift distribution is obtained from the usual cosmological relation

$$\frac{dN}{dSdz} = \int_{M0}^{M1} \phi_{BL}(M,z) \frac{dV}{dz} dM \quad , \quad (5)$$

where $\Phi_{BL}(M,z)$ is the beamed luminosity function, $S$ the surface area and $dV/dz$ the cosmological volume element per unit redshift. We compute the z dependence of the luminosity function from

$$\Phi_{BL}(M,z) = \Phi_{BL}(M, z=0)(1+z)^b \quad . \quad (6)$$

determined by Morris et al. (1991). Ajello et al. (2014) use this dependence multiplied by $e^{z/\xi}$, which we neglect to minimize the number of free parameters. In practice the effect of the $e^{z/\xi}$ is very small for it would only slightly sharpen the peak in Figure 1. We use $b = -4$ in Equation 6, in agreement with the redshift dependences in figure 11 in Ajello et al. (2014). For a proper estimate of the very important normalization parameter $K$ with Equation 4, we use the same cosmological parameters ($H_0 = 70$ *Km/sec/Mpc* and $\Omega = 0.7$) in Ross et al. (2012) since we use their QSO luminosity function. We then obtain the redshift distribution given by the dashed line in Figure 1. The histogram in Figure 1 gives the observed redshift distributions of the BLLs in figure 9 of Plotkin et al. (2008). Figure 1 shows that the numerical simulation fits reasonably well the observed redshift distribution, particularly so if we make the reasonable assumption that the uncertainty in the data can be quantified by standard deviations that are obtained from the square roots of the numbers in the simulations.

## 3. DISCUSSION AND CONCLUSION

We have quantitatively evaluated the hypothesis that BL Lacertae objects are the remnants of Quasi Stellar Objects. This hypothesis is based on the fact that the comoving space density of BLLs shows a strong increase with cosmological time (negative evolution with redshift). We start from the beamed luminosity function (Equation 2) from Urry & Shafer (1984) who consider the effects of relativistic beaming on a power law luminosity function. All of the values of the free parameters in our simulations are compatible with usual assumptions and within ranges constrained by observations. The main free parameter $K$ in Equation (3) comes from Equation (4) which makes our fundamental assumption that all QSOs are capable of generating relativistic jets, but that the jets only emerge and are detected in their old age to generate BLLs. The QSO remnant is detected as a BLL only if the angle between the beam and the line of sight is small ($\theta < 7°$).

Figure 1 shows an excellent agreement between our numerical simulations and the redshift distribution of BLLs from the SDSS survey (Plotkin et al. 2008). The advantage of using the SDSS BLLs is that it allows a comparison with data from the SDSS survey for both BLLs and QSOs, which is fundamental in getting the most important parameter $K$ in our simulations from Equation 4. Note also that the SDSS BLL redshift distribution that we use in Figure 1 is in agreement with other spectroscopic surveys (Ackerman et al. 2011, Massaro et al. 2009).

We used a model as simple as possible to minimize the number of free parameters. We assume that all sources have the same value of relativistic factor $\gamma$, use a power-law luminosity function and the simple *z* dependence for the evolution of the luminosity function in Equation 6. The number of observed BLLs is small, as can be seen in the histogram in Figure 1, so that the uncertainties in the observations are large. Consequently this does not justify the use of additional free parameters in our models or the use of more complex luminosity functions.

One must consider selection effects, particularly coming from the fact that BLLs have weak emission and absorption lines that are often below detection. These selection effects are discussed in Ajello et al. (2014), Giommi et al. (2012) and Giommi et al. (2013). The total sample of candidates from Plotkin et al. (2008) has 501 objects. The histogram in Figure 1 only uses the 256 objects that have reliable or lower limit redshifts in Plotkin et al. (2008) so that almost half of the candidates have no redshift. Let us first note that, as stated in the caption of figure 9 in Plotkin et al (2008) all redshifts with *z* <*1.1* , that are the main contributors to our simulations, are derived from host galaxy features, so that most of the redshifts in the histogram in Figure 1 are reliable. We must however consider the worst-case scenario that the majority of the 245 BLLs without redshift are at *z* > 0.6 and therefore weaken the negative evolution.

We can verify the validity of a no evolution hypothesis by comparing a numerical simulation for the case where there is no evolution whatsoever ($b = 0.0$), given by the dashed line in Figure 2, to a worst-case redshift distribution given by the histogram in Figure 2, where we assume that the 245 objects without redshift have a uniform redshift distribution for $0.6 < z < 3$ and add them to the 256 objects with redshifts. We see that the simulations poorly fit the observations. We arbitrarily normalize the theoretical counts to best fit the peak of the histogram. Note that the fact that, unlike in Figure 1, the theoretical counts for $z > 2.0$ are large does not affect the discussion because, unlike in Figure 1, we do not assume that most BL Lacs are evolved QSOs with *1.82 < z < 2.20*.

It is impossible to accurately estimate the effect of a better fit and better redshift distribution since we do not know the redshifts of the 245 objects without redshifts. We can however make a simple quantitative evaluation of a hypothetical better fit to a better redshift distribution for the 245 objects by considering that the numerical simulation predict that the number of objects at $z > 0.6$ should be more than three times the number at $z < 0.6$. Consequently, only 125 BLLs of the 501 sample should be at $z < 0.6$ and 376 at $z > 0.6$, while 201 BLLs with redshifts are at $z < 0.6$ and, assuming that the 234 without redshift are at $z > 0.6$, 300 are at $z > 0.6$. By computing the standard deviation $\sigma$ from the square root of the counts, we obtain $\sigma = 11.2$ for the counts of BLLs with $z < 0.6$ so that the counts are $6.8\sigma$ away, and $\sigma = 19.4$ for the counts of BLLs with $z>0.6$ so that the counts are at $3.9\sigma$ away. Consequently, this unlikely worst-case scenario can only weaken evolution, so that $b > -4$ in Equation 6, but not eliminate it.

In our previous discussion we consider together the different types of BLLs, while Ajello et al (2014) show that the evolution depends on the type. The luminosity functions in figure 11 in Ajello et al. (2014), as stated in their figure caption, show that the negative evolution is caused by low-luminosity HSPs. The ISP and LSP high-luminosity BLLs are the objects that predominate at high z and undergo positive evolution. Consequently we can make the hypothesis that only HSP BLLs are beamed QSO remnants.

Our analysis, like all studies of BLLs, has weaknesses that come from the small numbers of objects, difficulties measuring redshifts and the fact that objects identified as BLLs contain a mixture of different objects classified as BLLs only because they have the type of spectrum that is generated by a relativistic jet. These weaknesses, discussed in the previous paragraphs, should not affect greatly our main conclusion that HSP BLLs are QSO remnants. This is because Ajello et al. (2014) clearly show that HSPs are the only BLLs that undergo negative evolution and, furthermore, the redshift problem is far less important for HSPs which have the most reliable redshifts and dominate at low redshifts. In particular we must note that the redshift bias in spectroscopic surveys discussed by Ajello et al. (2014) affects little our use of SDSS BLLs from Plotkin et al.

(2008) to obtain, from Equation 4, the most important *K* parameter in Equation 3, since we use counts at z < 0.6, where HSP dominate and also where the redshifts are the most reliable and consequently we can assume that most BLLs at z <0.6 in Plotkin et al. (2008) are HSPs.

In conclusion, our numerical simulations show that the hypothesis that HSP BLLs are relativistically beamed QSO is in agreement with the observed redshift distribution of BLLs and that it should be taken seriously.

Finally, if the hypothesis is correct the knowledge of the local space density BLLs and of quasars at high redshifts should give a powerful cosmological test (Borra 1983).

**Acknowledgments**

This research was supported by the Natural Sciences and Engineering Research council of Canada.

**FIGURE CAPTIONS**

Figure 1: The dashed line gives the redshift distribution obtained from the model described in section 2 assuming that the luminosity function evolves with cosmic time proportionally to $(1+z)^b$ with $b = -4$ (negative evolution). The counts are quantized in redshift bins having width $\Delta z = 0.2$. The histogram shows the observed redshift distributions of the SDSS BL Lac objects in figure 9 of Plotkin et al. 2008.

Figure 2: It gives the redshift distribution, obtained from the model described in section 2, assuming that the luminosity function does not evolve with cosmic time ($b = 0$). The counts are quantized in redshift bins having width $\Delta z = 0.2$. The histogram assumes that the 245 objects without redshift have a uniform redshift distribution for $0.6 < z < 3$ and adds them to the 256 objects with redshifts

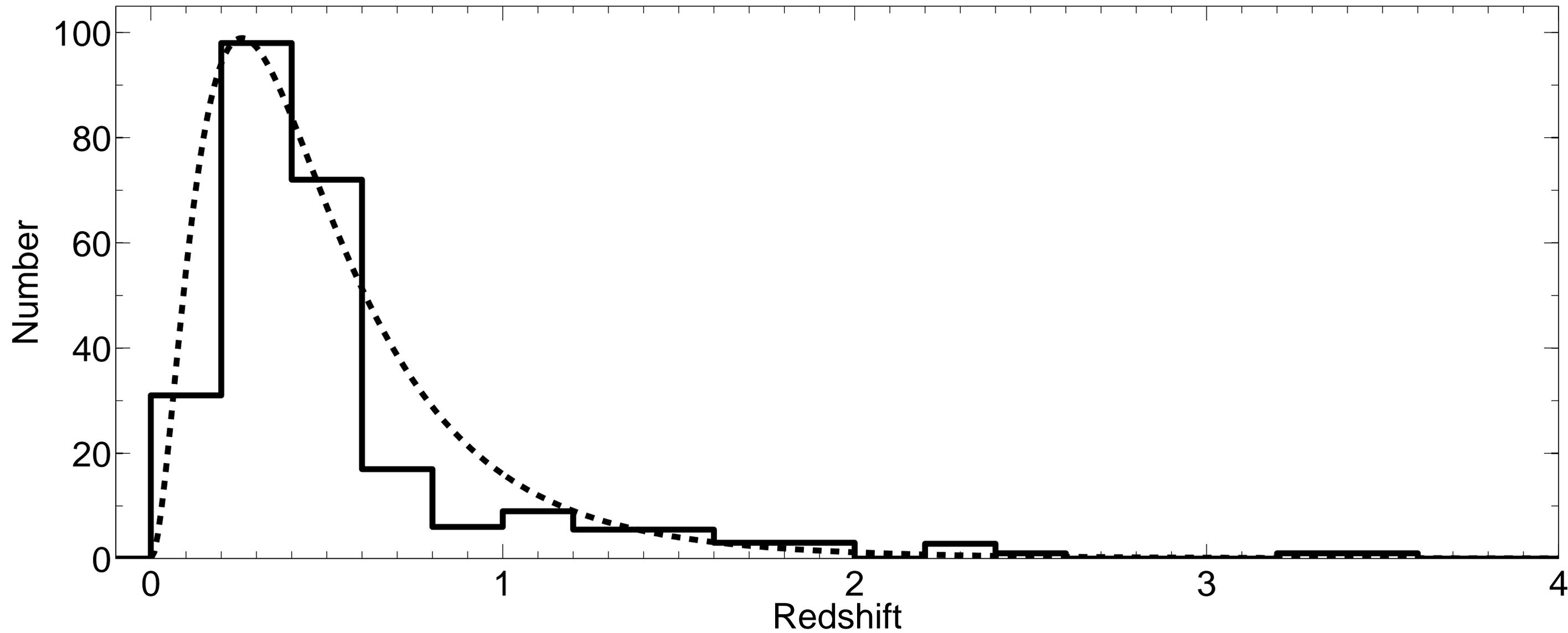

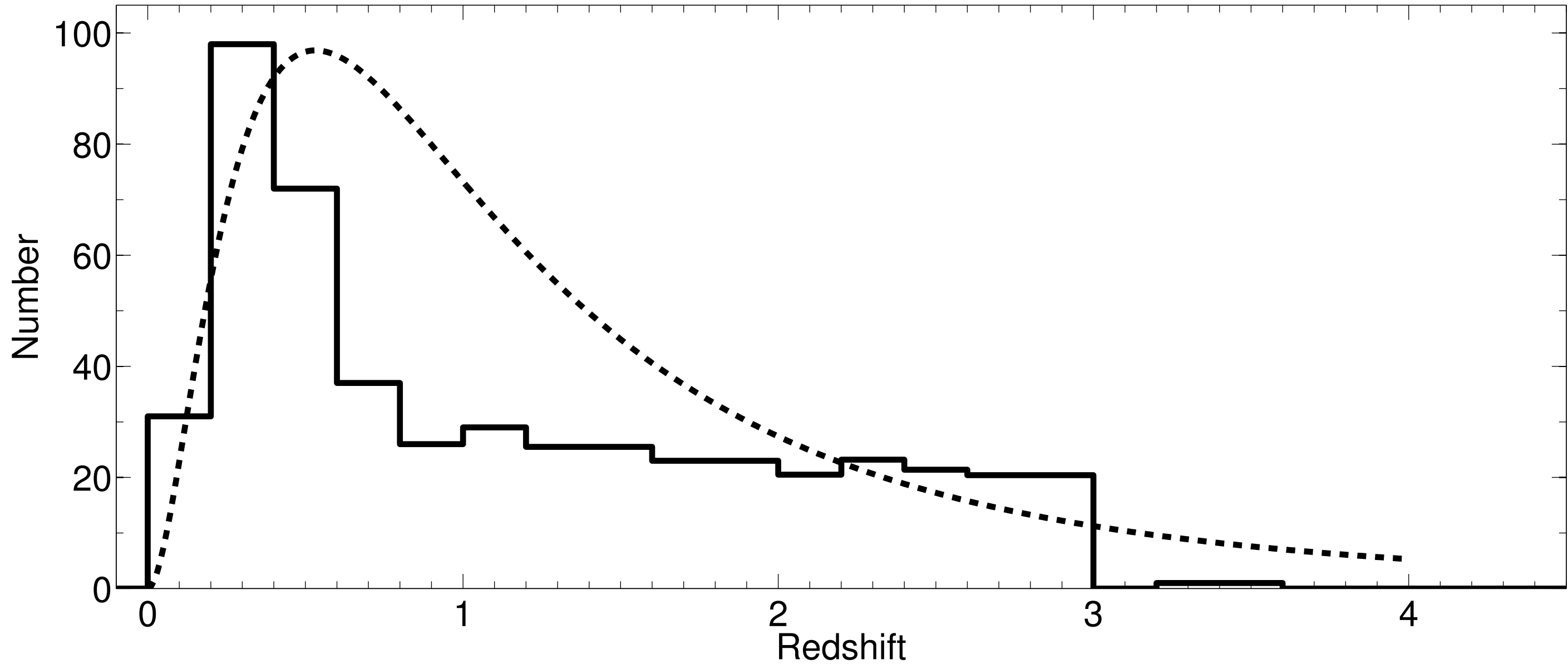